# Towards calibration-free Mach-Zehnder switches on silicon

LIJIA SONG,[1] TANGNAN CHEN,[1] WEIXI LIU,[1] HONGXUAN LIU,[1] YINGYING PENG,[1] ZEJIE YU,[1] HUAN LI,[1] YAOCHENG SHI[1], AND DAOXIN DAI [1,2] [*]

[1]*State Key Laboratory for Modern Optical Instrumentation, College of Optical Science and Engineering, International Research Center for Advanced Photonics, Zhejiang University, Zijingang Campus, Hangzhou 310058, China.*
[2]*Ningbo Research Institute, Zhejiang University, Ningbo 315100, China.*
*\* dxdai@zju.edu.cn*

**Abstract:** Silicon photonic Mach-Zehnder switches (MZSs) have been extensively investigated as a promising candidate for practical optical interconnects. However, conventional 2×2 MZSs are usually prone to the size variations of the arm waveguides due to imperfect fabrication, resulting in considerable random phase imbalance between the two arms, thereby imposing significant challenges for further scaling up N×N MZSs. Here we propose a novel design towards calibration-free 2×2 and N×N MZSs, employing optimally widened arm waveguides, enabled by novel compact tapered Euler S-bends with incorporated mode filters. With standard 180-nm CMOS foundry processes, more than thirty 2×2 MZSs and one 4×4 Benes MZS with the new design are fabricated and characterized. Compared with their conventional counterparts with 0.45-µm-wide arm waveguides, the present 2×2 MZSs exhibit ~370-fold reduction in the random phase imbalance. The measured extinction ratios of the present 2×2 and 4×4 MZSs operating in the all-cross state are ~30 dB and ~20 dB across the wavelength range of ~60 nm, respectively, even without any calibrations. This work paves the way towards calibration-free large-scale N×N silicon photonic MZSs.

## 1. Introduction

Mach-Zehnder interferometers (MZIs) have been recognized as an indispensable fundamental element in various optical systems due to their great versatility for diverse applications [1-5]. In the past decades, on-chip MZIs have been developed with diverse material systems and widely used as one of the most essential components in photonic integrated circuits (PICs) [6-14]. Among them, silicon MZIs are becoming increasingly attractive because silicon photonics features ultrahigh integration density as well as excellent CMOS (complementary metal-oxide semiconductor) compatibility. Silicon MZIs have been developed successfully for realizing variable optical couplers [15, 16], optical modulators [17], optical filters [18-21], variable optical attenuators [22], optical sensors [23], and optical switches[24-26]. Among them, Mach-Zehnder switches (MZSs) are one of the most representative functional elements and have been investigated for decades. In particular, thermo-optic (TO) MZSs feature excellent performances and design/fabrication simplicity [27-29], compared to their electro-optic (EO) counterparts based on carrier injection/depletion[24], and hence have been extensively investigated as a promising candidate for practical optical interconnects, such as optical burst switching (OBS) in the high-speed optical internet backbone.

Beyond a single 2×2 MZS, it is also very important to achieve N×N MZSs consisting of large-scale networks of 2×2 MZSs in cascade. For example, 16×16 and 32×32 MZSs have been realized on silicon in recent years[30-32]. In this case, the total number of elementary 2×2 MZSs scales up rapidly with the port count N, imposing increasingly stringent requirements on the performance of the elementary 2×2 MZSs. A typical 2×2 MZS is composed of two 2×2 3-dB couplers and two symmetric arms, designed for low excess losses and high extinction ratios. However, conventional 2×2 MZSs are usually prone to the random size variations of the arm waveguides due to the imperfect fabrication with the state-of-the-art CMOS foundry processes, resulting in considerable accumulated random phase errors and unpredictable phase imbalance between the two arms. In this case, such random phase imbalance must be calibrated and compensated meticulously for all the 2×2 MZSs one by one in a large-scale N×N MZS. Therefore, a large number of additional power taps as well as power monitors are often required for all or part of the 2×2 MZS elements, so that the optimal electrical power for their cross- and bar-states can be individually determined by monitoring the corresponding tapped power. However, this inevitably introduces some significant excess losses. Furthermore, it also entails additional on-chip feedback control schemes and sophisticated characterization procedures, which significantly complicates the layout design and greatly increases the chip footprint as well as the chip management complexity. Besides, it also consumes extra heating power for both cross- and bar-states. Therefore, it becomes very challenging to scale up N×N MZSs further.

Recently, in [33] we proposed and demonstrated a new design of 2×2 MZSs with lowered random phase errors for the first time by widening the straight phase-shifter waveguides, which effectively reduced the random phase imbalance, compared to the case of using conventional 0.45-µm-wide singlemode phase shifters. The fabrication tolerance is thus improved and the power consumption for compensating the phase-imbalance is considerably reduced. Here we further propose a novel design towards *calibration-free* 2×2 and N×N MZSs that can be mass-manufactured in state-of-the-art silicon photonics foundries. The effective methodology to minimize the size-variation sensitivity of the elementary 2×2 MZS is to judiciously widen the entire MZI arm waveguides, including not only the straight phase shifters but also the S-bends, the tapers, as well as the input/output waveguides of the 2×2 multimode-interference (MMI) couplers. Specifically, the present 2×2 MZS is designed and implemented by introducing novel tapered Euler S-bends (TES-bends) with a widened core width. Furthermore, bent asymmetric directional couplers (ADCs) are incorporated into the TES-bends to filter out residual higher-order modes at the TES-bend entrance. With a standard 180-nm CMOS foundry process, more than thirty 2×2 MZSs with the proposed new design on eleven silicon photonic chips were fabricated and characterized. Compared with those conventional 2×2 MZSs with 0.45-µm-wide singlemode phase shifters, the present 2×2 MZSs exhibit ~370-fold reduction in the random phase imbalance. This validates the improved fabrication tolerance and results in considerable reduction of the power consumption for the phase-imbalance compensation. Furthermore, a 4×4 MZS with Benes network topology is also fabricated with the same foundry process and characterized experimentally. The measured extinction ratios of the 2×2 MZS and the 4×4 MZS are ~30 dB and ~20 dB across a broad wavelength range of ~60 nm, respectively, even without any calibrations. This work paves the way towards *calibration-free* large-scale N×N silicon photonic MZSs. The proposed methodology for suppressing the random phase imbalance can be generalized for analog MZI elements and other essential phase-sensitive photonic integrated devices as well, such as micro-rings and arrayed-waveguide gratings (AWGs).

## 2. Design of the Elementary 2×2 MZS

Figure 1(a) shows the schematic configuration of the proposed elementary 2×2 MZS, which consists of two identical 2×2 MMI couplers and two arm waveguides, and Fig. 1(b) shows the cross-section of the phase-shifter with a TiN micro-heater on top. Each arm waveguide is composed of a phase shifter, two adiabatic tapers, two TES-bends, and the straight sections connecting to the input/output ports of the MMI couplers. Here the MMI coupler is designed with a 2.4-µm-wide and 20.5-µm-long MMI section, while the width of the input/output waveguides for the MMI coupler is chosen as 0.9 µm. The TES-bends and the widened phase-shifters are connected through a ~10-µm long adiabatic tapers (Supplement 1).

The S-bends and the phase-shifters of conventional MZSs are usually singlemode, and thus their effective indices depend sensitively on the random variation of the core width due to the fabrication errors. Consequently, notable accumulated phase imbalance is usually observed between the two MZS arms due to the random difference in their core widths. A promising solution to this problem is to introduce widened phase-shifters, as proposed in our previous work [33], where the straight section in the phase shifter are designed to be as wide as 2 μm. One might notice that the S-bends in the MZS arms in [33] are still as narrow as 450 nm to be singlemode, in order to avoid higher-order mode excitation. Such singlemode S-bends become the dominant contributor to the phase imbalance in the MZS arms.

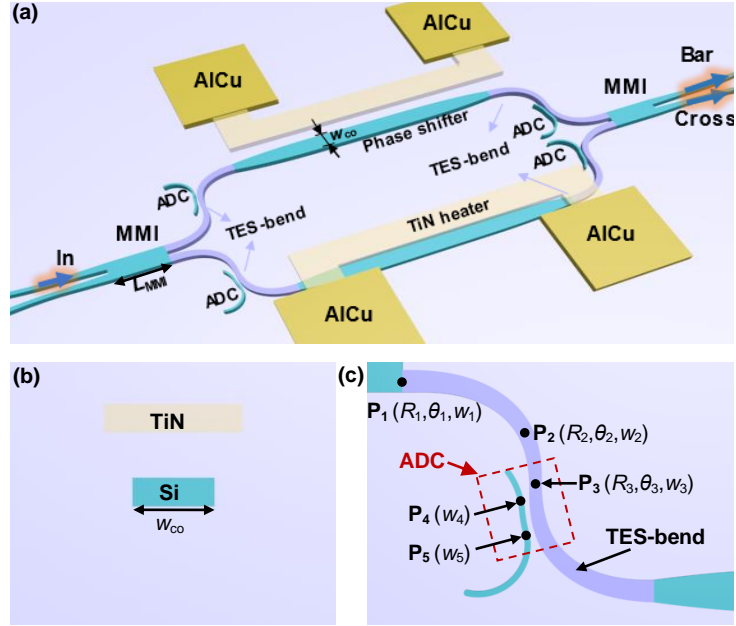

Fig. 1. Schematic configurations of the proposed elementary 2×2 MZS (to be calibration-free). (a) Overview; (b) Cross-section of the phase shifter; (c) TES-bend and the bent-ADC mode filter.

In contrast, in this paper it is the first time to incorporate specially-designed multimode S-bends into MZSs, whose effective indices are much less dependent on the core-width variation. As a result, the random phase imbalances can be greatly reduced. The challenge for such multimode S-bends in an MZS lies in the suppression of higher-order modes when light propagates in the arm waveguides. More specifically, the multimode S-bends should be designed with two key strategies to enable low-loss and low-crosstalk propagation of the fundamental mode. *One* is to quickly fan out the two waveguides connected with the MMI couplers to avoid any undesired evanescent coupling and thermal crosstalk between them. The *other* is to filter out the residual higher-order modes, especially the dominant $TE_1$ mode, which is slightly excited with a power ratio of $<-15$ dB due to imperfect self-imaging at the junction of the input MMI coupler (Supplement 1). Furthermore, the S-bends should be as compact as possible to reduce the accumulated phase imbalance.

Here a specially tapered Euler S-bend (TES-bend) with an incorporated bent-ADC mode filter is introduced, as shown in Fig. 1(c). The TES-bend consists of two identical 90° Euler-bends, whose curvature radius linearly decreases, with respect to the curve length $L$, from the radius $R_1$ at the end point $P_1$ (i.e., $\theta = \theta_1$) to the radius $R_2$ at the knee point $P_2$ (i.e., $\theta = \theta_2$), and then linearly increases to the radius $R_3$ at the deflection point $P_3$ (i.e., $\theta = \theta_3$), as shown in Fig.

2(a). Meanwhile, its width varies linearly with respect to the local angle $\theta$, between the width $w_1$ at the end points $P_1$, the width $w_2$ at the knee points $P_2$, and the width $w_3$ at the deflection point $P_3$, as shown in Fig. 2(b). Here the core width $w_1$ at the end point $P_1$ ($\theta_1 = 0°$) is chosen as $w_1 = 0.9$ µm to match the MMI coupler input/output ports. Accordingly, the curvature radius $R_1$ is chosen as $R_1 = 12$ µm to simultaneously minimize the footprint and the mode mismatch between the MMI coupler input/output waveguides and the TES-bend. The core width $w_3$ and the bending radius $R_3$ at the deflection point ($\theta_3 = 90°$) are chosen as $w_3 = 0.5$ µm and $R_3 = 10$ µm, respectively, so as to simultaneously minimize the footprint and the mode mismatch between those two sections with opposite curvatures. The other parameters at the knee points are optimally chosen as $\theta_2 = 60°$, $w_2 = 0.67$ µm, and $R_2 = 3$ µm. With this design, the footprint of the TES-bend is as small as ~9×9 µm².

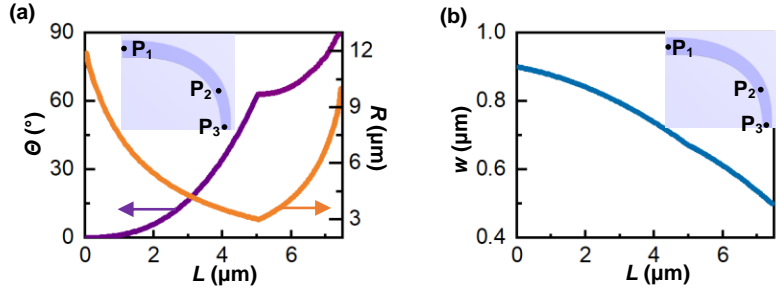

Fig. 2. The local angle $\theta$ and the curvature radius $R$ (a), and the core width $w$ (b) as functions of the curve length $L$ for the first half of the TES-bend.

Figure 3(a) shows the simulated light propagation in the designed TES-bend, where the propagation loss for the $TE_0$ mode is negligibly low (less than 0.07 dB), while that for the $TE_1$ mode is much higher (i.e., 1.97 dB) at 1550 nm. To further filter out the residual $TE_1$ mode, a bent-ADC mode filter is incorporated into the second half of the TES-bend, without increasing the TES-bend footprint, as shown in Fig. 1(c). The bent-ADC mode filter consists of a coupling region where a narrow waveguide is introduced on the convex side near the deflection point of the TES-bend. There is a 0.2-µm-wide gap in the 1.3-µm-long coupling region, and the width of the narrow waveguide varies from 0.24 µm to 0.26 µm. In this way, the $TE_1$ mode in the TES-bend can be coupled out to the $TE_0$ mode in the narrow waveguide, leading to decent mode-filtering within a highly compact footprint. Figure 3(b) shows the simulation results for light propagation of the $TE_0$ and $TE_1$ modes with the incorporation of the bent-ADC mode filter, which evidently suppresses the $TE_1$-mode transmission by 12-17 dB, meanwhile introduces negligible transmission loss for the $TE_0$ mode.

Figure 3(c) The wavelength-dependent transmissions are also calculated and the powers of the $TE_0$ and $TE_1$ modes at the cross and through ports. The simulated light propagation in the MMI coupler connected with the TES-bends are shown in Fig. 3(c). It exhibits excellent performances with low non-uniformity of <0.08 dB, low excess losses of ~0.26 dB and well-suppressed higher-order-mode excitation of <−25 dB in the wavelength range from 1520 nm to 1580 nm. When operating at the central wavelength of 1550 nm, the non-uniformity is about 0.06 dB, low excess losses is ~0.13 dB and the well-suppressed higher-order-mode excitation is <−28 dB. Finally, for the entire MZS consisting of the designed MMI couplers and TES-bends, the transmissions at the cross/bar ports are calculated with the finite-difference time-domain (FDTD) method, as shown in Fig. 3(d). It exhibits excellent performances with low excess losses of < 0.58 dB and high extinction ratios of ~33 dB, respectively from 1520 nm to 1580 nm. Note that the excess loss increases slightly as the wavelength increases, which is due to the excess loss from the MMI couplers and can be reduced possibly with further optimization.

Figure 4(a) shows the calculated accumulated phase imbalance for the present MZS design as the mean width-difference $\delta w$ varies. In order to provide a quantitative comparison, the result for the conventional MZS is also given. As shown in Fig. 4(a), the designs with the present TES-bends as well as the widened phase-shifter with $w_{co} \geq 2$ µm exhibit ~10-fold lower phase imbalance than the conventional design with $w_{co}=0.45$ µm. Furthermore, the itemized phase imbalance [33] are analyzed and shown in Fig 4(b). It can be seen that the present TES-bend has a lowered phase error of 0.0064 π/nm, which is ~22-fold improvement over the conventional 450-nm-wide uniform S-bend. Note that the dominant contributor to the total phase imbalance is still the TES-bend for $\geq$2-µm-wide phase-shifters (Supplement 1).

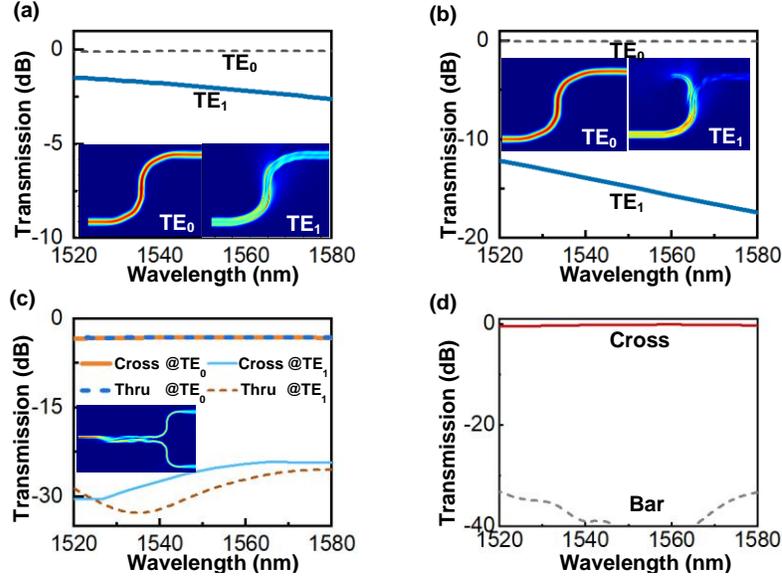

Fig. 3. Simulation results for the designed 2×2 MZS. Calculated transmissions of the designed TES-bend without (a) and with (b) the bent-ADC mode filter, respectively. Insets: simulated light propagations when the $TE_0$ and $TE_1$ modes are launched respectively; (c) Calculated transmissions of the MMI coupler connected with the TES-bends. Inset: simulated light propagation for the launched $TE_0$ mode; (d) Calculated transmissions at the cross and bar ports of the designed MZS in the OFF state.

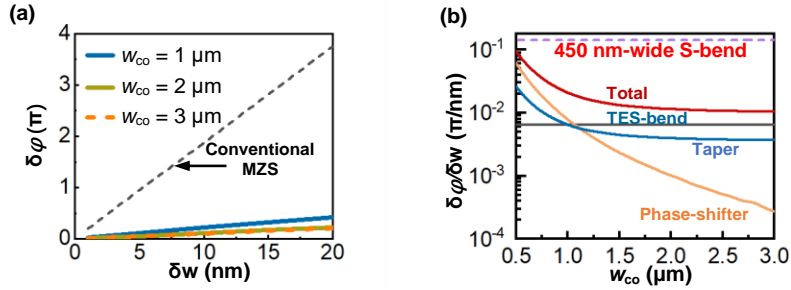

Fig. 4. (a) Calculated total phase imbalance for the new MZS consisting of TES-bends as well as arm-waveguides with different core-widths of 1, 2, and 3 µm; Here the mean width difference δw varies from 1 nm to 20 nm; The result for conventional MZSs with 450-nm-wide arm-waveguides is also given; (b) The itemized phase imbalance as the core width $w_{co}$ varies. The results for the conventional MZS with 450-nm-wide S-bends are also included for comparison.

## 3. Fabrication and Measurement

Here the MZSs have been fabricated on silicon-on-insulator (SOI) wafers with a 220-nm-thick top silicon layer and a 2-µm-thick buried oxide (BOX) layer, using standard 180-nm CMOS foundry processes, as shown in Fig. 5(a). Here TiN micro-heaters are on top of both phase shifters and grating couplers are used for convenient and efficient chip-fiber coupling.

Figures 5(b) and 5(c) show the measurement results for one of the representative MZSs with TES-bends and 2-µm-wide phase-shifters (Design A). From Fig. 5(b), which shows the measured transmissions at the cross/bar ports at the central wavelength when sweeping the heating power from 0 to 80 mW, the phase imbalance is very small and the corresponding heating power for the compensation is ~0.66 mW only. The bar port has an extinction ratio as high as 32 dB even without any heating power (i.e., $Q$=0) for phase compensation. From Fig. 5(c), which shows the transmissions at the cross/bar ports of the MZS operating at the ON/OFF states (i.e., $Q$=0, and 34 mW), the extinction ratios are >30 dB in the wavelength range from 1520 nm to 1580 nm. These results verify the excellent performance of the present MZS without any additional power consumption for the phase-imbalance compensation, paving the way towards to calibration-free MZSs.

Figure 5(d) provides a summary for the calibrated phase imbalances for all the MZSs from 11 chips in the same fabrication batch. These MZSs have three different types of designs. The first one is the Design A described above, the second one is Design B with the present TES-bends but 1-µm-wide phase-shifters, and the third one is Design C of the conventional MZS whose S-bends and phase-shifters are 0.45-µm wide. As shown in Fig. 5(d), the phase-imbalance compensation needed for the conventional design (Design C) is $(0.70\pm0.50)\pi$, with a large standard deviation, which indicates that Design C leads to highly random phase imbalances. Consequently, sophisticated and meticulous calibration and compensation are required for each MZS, which consumes additional power.

In contrast, Designs A and B exhibit a significant improvement compared to the conventional design (Design C). As shown in Fig. 5(d), the phase-imbalance compensations for them are $(0.016\pm0.045)\pi$ and $(0.0019\pm0.077)\pi$, respectively. Compared with the conventional design, the new designs exhibit ~370-fold reduction in the random phase imbalance $\delta\varphi$. Therefore, it is feasible to achieve superior performance with high extinction ratios and low excess losses for Designs A & B even without any calibration for the OFF state. This promises to greatly simplify the calibration particularly for N×N MZSs with a large number of 2×2 MZSs, as well as to reduce the power consumption significantly for phase-imbalance compensation.

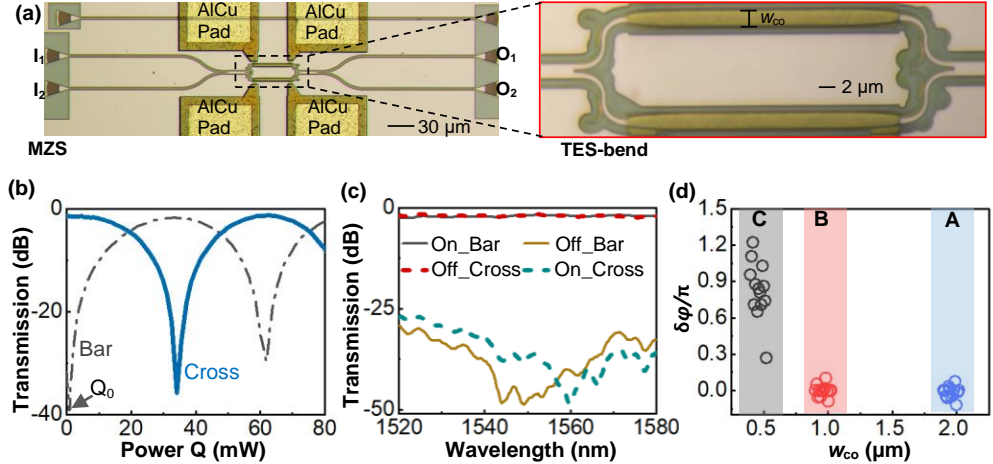

Fig. 5. (a) Optical microscope image of the fabricated 2×2 MZS; (b) Measured transmissions at the cross/bar ports for the central wavelength when sweeping the heating power $Q$ from 0 to 80 mW; (c) Measured transmissions at the cross/bar ports of the present MZS operating at the ON/OFF states (i.e., $Q$=0, and 34 mW); (d) Summary of the measured phase imbalances for all the MZSs (Designs A, B, and C) from 11 chips in the same fabrication batch. Design A is with the present TES-bends and 2-µm-wide phase-shifters; Design B is with the present TES-bends and 1-µm-wide phase-shifters; Design C is with the conventional 0.45-µm-wide S-bends/phase-shifters.

The present 2×2 MZS with Design A is used further for the realization of a 4×4 MZS with Benes network, which consists of six 2×2 MZSs cascaded in three stages, as shown in Fig. 6(a). Here we only present the measured transmission spectra from the most representative switching states, i.e., all-cross, all-bar, and the six single-bar switching states (see Supplement 1). When the 4 × 4 MZS works in the all-cross state, the signals launched from input ports $I_1$, $I_2$, $I_3$, and $I_4$ are routed to output ports $O_3$, $O_4$, $O_1$, and $O_2$, respectively. Figures 6(b)-(e) show the measured transmissions $T_{ij}$ at the output port $O_j$ from the input port $I_i$ when $i$=1, 2, 3, and 4, respectively, with no calibration for all the six MZSs (i.e., the powers applied to the six micro-heaters are zero). The extinction ratios for all the ports are ~20 dB across the 60-nm bandwidth. Figures 6(f)-(i) shows the measured all-bar transmissions $T_{ij}$ defined above. When the 4 × 4 MZS works in the all-bar state, the signals launched from input ports $I_1$, $I_2$, $I_3$, and $I_4$ are routed to output ports $O_3$, $O_4$, $O_1$, and $O_2$, respectively. The measured all-bar transmissions of the 4 × 4 MZS exhibit excellent performances, similar to the cases with all-cross states.

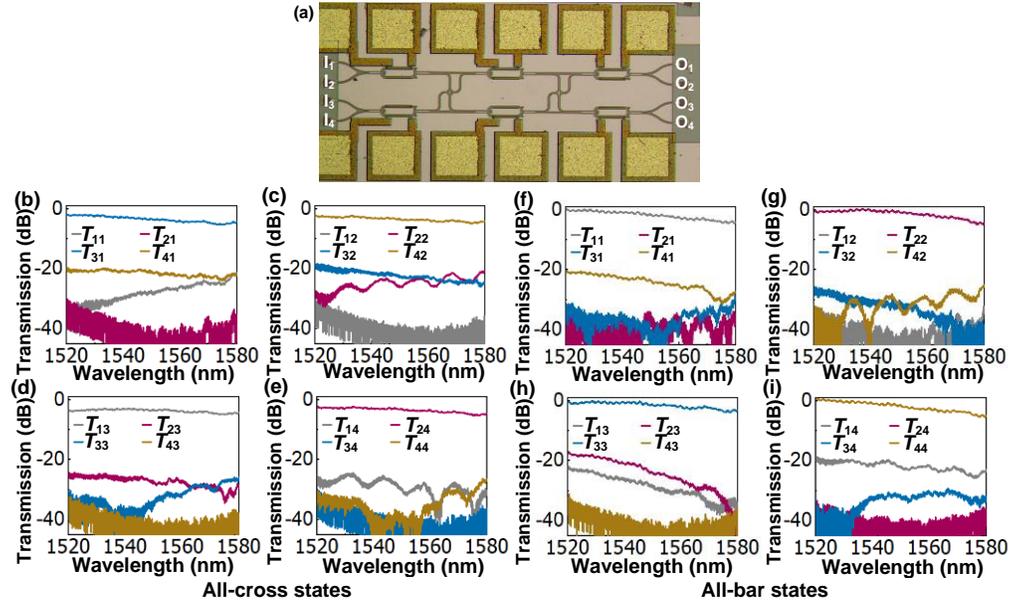

Fig.6. Optical microscope image (a), and the measured all-cross transmissions $T_{ij}$ at the output port $O_j$ with the port $I_i$ importing when $i$=1, 2, 3, and 4, respectively, with no calibration for all the six MZSs. (b)-(e) and the measured all-bar transmissions $T_{ij}$ (f)-(i) which the signals launched from input ports $I_1$, $I_2$, $I_3$, and $I_4$ are routed to output ports $O_1$, $O_2$, $O_3$, and $O_4$, respectively.

The 4×4 MZS was further used to demonstrate high-bit-rate data routing. In order to characterize the signal integrity degradation due to the MZS crosstalk, eye-diagrams at any output port $O_j$ should be recorded when all four data channels are launched into the input ports $I_i$ ($i$=1, 2, 3, and 4). However, due to the lack of concurrent data generators in the lab, we synthesize the eye-diagrams at the output port $O_j$ from the measured transmissions $T_{ij}$ by reasonably assuming the four inputs are incoherent. In our experiments, the 30-Gbps Non-Return-to-Zero (NRZ) data were launched into the input ports one by one, and the transmissions $T_{ij}$ from input port $I_i$ to output port $O_j$ were measured and recorded. The eye-diagram $T_j$ at the output port $O_j$ is synthesized by summing the transmissions $T_{ij}$ ($i$ = 1, 2, 3, and 4), i.e., $T_j = \sum_i T_{ij}$. Figures 7(a)-(d) show the synthesized eye-diagrams for ports $O_1$, $O_2$, $O_3$, and $O_4$, respectively, exhibiting open eye-diagrams with high signal-to-noise ratios, which validates the high-bit-rate data routing of the present 4 × 4 MZS in the all-cross states.

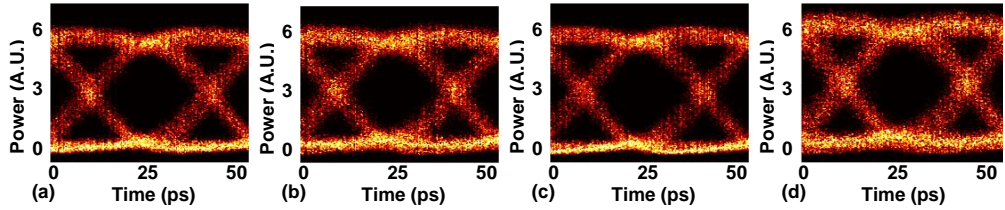

Fig.7. Synthesized eye-diagrams at port $O_1$ (a), $O_2$ (b), $O_3$ (c), and $O_4$ (d) of the present 4 × 4 MZS in the all-cross states. Here the bit rate is 30 Gbps.

## 4. Conclusions

In conclusion, we have proposed a novel design towards calibration-free 2×2 and N×N MZSs that can be mass-manufactured in state-of-the-art silicon photonics foundries, by judiciously widening the MZI arm waveguides. Specifically, the present 2×2 MZS is designed and implemented by introducing novel TES-bends with a widened core width and incorporated bent-ADC mode filters. With a standard 180-nm CMOS foundry process, more than thirty 2×2 MZSs and one 4×4 Benes MZS with the new design have been fabricated and characterized. Compared with those conventional 2×2 MZS with 0.45-µm-wide singlemode phase-shifters, the present 2×2 MZSs with 2-µm-wide phase-shifters exhibit a small random phase imbalance of $(0.016±0.045)π$, which is a significant improvement compared to the conventional design with 0.45-µm-wide singlemode phase shifters. The fabricated 2×2 and 4×4 MZSs have high extinction ratios of ~30 dB and ~20 dB across a ~60-nm-wide wavelength range, respectively, even without any calibrations. These results pave the way towards large-scale calibration-free N×N silicon photonic MZSs. The proposed methodology for suppressing the random phase imbalance can be generalized for analog MZI elements and other essential phase-sensitive integrated photonic elements as well, such as micro-rings and AWGs.

**Funding.** This work was supported by National Major Research and Development Program (2019YFB2203603), Zhejiang Provincial Major Research and Development Program (No. 2021C01199), National Science Fund for Distinguished Young Scholars (61725503), National Natural Science Foundation of China (NSFC) (91950205, 61961146003), Zhejiang Provincial Natural Science Foundation (LZ18F050001, LD19F050001), and the Fundamental Research Funds for the Central Universities.

**Disclosures.** The authors declare no conflicts of interest.

**Data Availability.** Data underlying the results presented in this paper are not publicly available at this time but may be obtained from the authors upon reasonable request.

**Supplemental document.** See Supplement 1 for supporting content.

**References.**
1. Y. Li and L. Tong, "Mach-Zehnder interferometers assembled with optical microfibers or nanofibers," Opt. Lett. **33**, 303-305 (2008).
2. D. A. B. Miller, "Perfect optics with imperfect components," Optica **2**, 747-750 (2015).
3. Y. Zhang, Q. Du, C. Wang, T. Fakhrul, S. Liu, L. Deng, D. Huang, P. Pintus, J. Bowers, C. A. Ross, J. Hu, and L. Bi, "Monolithic integration of broadband optical isolators for polarization-diverse silicon photonics," Optica **6**, 473-478 (2019).
4. S. Ghosh, S. Keyvaninia, Y. Shoji, W. Roy, T. Mizumoto, G. Roelkens, and R. Baets, "Compact Mach-Zehnder interferometer Ce:YIG/SOI optical isolators," IEEE Photonics Technology Letters **24**, 1653-1656 (2012).
5. C. Zhang, P. Morton, J. Khurgin, J. Peters, and J. Bowers, "Ultralinear heterogeneously integrated ring-assisted Mach–Zehnder interferometer modulator on silicon," Optica **3**, 1483 (2016).
6. A. M. Al-Hetar, A. B. Mohammad, A. S. M. Supa'at, and Z. A. Shamsan, "MMI-MZI Polymer Thermo-Optic Switch With a High Refractive Index Contrast," J. Lightwave Technol. **29**, 171-178 (2011).
7. M. He, M. Xu, Y. Ren, J. Jian, Z. Ruan, Y. Xu, S. Gao, S. Sun, X. Wen, L. Zhou, L. Liu, C. Guo, H. Chen, S. Yu, L. Liu, and X. Cai, "High-performance hybrid silicon and lithium niobate Mach–Zehnder modulators for 100 Gbit s −1 and beyond," Nature Photonics **13**, 1 (2019).

8. A. Ribeiro, A. Ruocco, L. Vanacker, and W. Bogaerts, "Demonstration of a 4×4-port universal linear circuit," Optica **3**, 1348-1357 (2016).
9. J. S. Barton, E. J. Skogen, M. L. Masanovic, S. P. Denbaars, and L. A. Coldren, "A widely tunable high-speed transmitter using an integrated SGDBR laser-semiconductor optical amplifier and Mach-Zehnder modulator," IEEE Journal of Selected Topics in Quantum Electronics **9**, 1113-1117 (2003).
10. R. Amin, R. Maiti, Y. Gui, C. Suer, M. Miscuglio, E. Heidari, R. T. Chen, H. Dalir, and V. J. Sorger, "Sub-wavelength GHz-fast broadband ITO Mach–Zehnder modulator on silicon photonics," Optica **7**, 333-335 (2020).
11. H. Wang, H. Chai, Z. Lv, Z. Zhang, L. Meng, X. Yang, and T. Yang, "Silicon photonic transceivers for application in data centers," Journal of Semiconductors **41**, 101301 (2020).
12. P. Bhasker, J. Norman, J. Bowers, and N. Dagli, "Intensity and Phase Modulators at 1.55 μm in GaAs/AlGaAs Layers Directly Grown on Silicon," J. Lightwave Technol. **36**, 1-1 (2018).
13. D. Korn, R. Palmer, H. Yu, P. Schindler, L. Alloatti, M. Baier, R. Schmogrow, W. Bogaerts, S. K. Selvaraja, G. Lepage, M. Pantouvaki, J. Wouters, P. Verheyen, J. Van Campenhout, B. Chen, R. Baets, P. P. Absil, R. Dinu, C. Koos, and J. Leuthold, "Silicon-organic hybrid (SOH) IQ modulator using the linear electro-optic effect for transmitting 16QAM at 112 Gbit/s," Opt. Express **21**, 13219-13227 (2013).
14. X. Yang, M. S. Nisar, W. Yuan, F. Zheng, L. Lu, J. Chen, and L. Zhou, "Phase change material enabled 2 × 2 silicon nonvolatile optical switch," Opt. Lett. **46**, 4224-4227 (2021).
15. W. R. Clements, P. C. Humphreys, B. J. Metcalf, W. S. Kolthammer, and I. A. Walmsley, "Optimal design for universal multiport interferometers," Optica **3**, 1460-1465 (2016).
16. M. Reck, A. Zeilinger, H. J. Bernstein, and P. Bertani, "Experimental realization of any discrete unitary operator," Physical Review Letters **73**, 58-61 (1994).
17. A. Liu, R. Jones, L. Liao, D. Samara-Rubio, D. Rubin, O. Cohen, R. Nicolaescu, and M. Paniccia, "A high-speed silicon optical modulator based on a metal–oxide–semiconductor capacitor," Nature **427**, 615-618 (2004).
18. R. A. Soref, F. De Leonardis, and V. M. N. Passaro, "Reconfigurable optical-microwave filter banks using thermo-optically tuned Bragg Mach-Zehnder devices," Opt. Express **26**, 14879-14893 (2018).
19. F. Horst, W. M. J. Green, S. Assefa, S. M. Shank, Y. A. Vlasov, and B. J. Offrein, "Cascaded Mach-Zehnder wavelength filters in silicon photonics for low loss and flat pass-band WDM (de-)multiplexing," Opt. Express **21**, 11652-11658 (2013).
20. M. E. Ganbold, H. Nagai, Y. Mori, K. Suzuki, H. Matsuura, K. Tanizawa, K. Ikeda, S. Namiki, H. Kawashima, and K. i. Sato, "A Large-Scale Optical Circuit Switch Using Fast Wavelength-Tunable and Bandwidth-Variable Filters," IEEE Photonics Technology Letters **30**, 1439-1442 (2018).
21. L. Shen, L. Lu, Z. Guo, L. Zhou, and J. Chen, "Silicon optical filters reconfigured from a 16 × 16 Benes switch matrix," Opt. Express **27**, 16945-16957 (2019).
22. Q. Wu, L. Zhou, X. Sun, H. Zhu, L. Lu, and J. Chen, "Silicon thermo-optic variable optical attenuators based on Mach–Zehnder interference structures," Optics Communications **341**, 69-73 (2015).
23. K. Misiakos, I. Raptis, E. Makarona, A. Botsialas, A. Salapatas, P. Oikonomou, A. Psarouli, P. S. Petrou, S. E. Kakabakos, K. Tukkiniemi, M. Sopanen, and G. Jobst, "All-silicon monolithic Mach-Zehnder interferometer as a refractive index and bio-chemical sensor," Opt. Express **22**, 26803-26813 (2014).
24. M. Yang, W. M. J. Green, S. Assefa, J. Van Campenhout, B. G. Lee, C. V. Jahnes, F. E. Doany, C. L. Schow, J. A. Kash, and Y. A. Vlasov, "Non-Blocking 4×4 Electro-Optic Silicon Switch for On-Chip Photonic Networks," Opt. Express **19**, 47-54 (2011).
25. N. Dupuis, A. V. Rylyakov, C. L. Schow, D. M. Kuchta, C. W. Baks, J. S. Orcutt, D. M. Gill, W. M. J. Green, and B. G. Lee, "Ultralow crosstalk nanosecond-scale nested 2 × 2 Mach–Zehnder silicon photonic switch," Opt. Lett. **41**, 3002-3005 (2016).
26. K. Suzuki, R. Konoike, G. Cong, K. Yamada, S. Namiki, H. Kawashima, and K. Ikeda, "Strictly Non-Blocking 8 × 8 Silicon Photonics Switch Operating in the O-Band," J. Lightwave Technol. **39**, 1096-1101 (2021).
27. S. Wang and D. Dai, "Polarization-insensitive 2×2 thermo-optic Mach-Zehnder switch on silicon," Opt. Lett. **43**, 2531-2534 (2018).
28. F. Duan, K. Chen, D. Chen, and Y. Yu, "Low-power and high-speed 2 × 2 thermo-optic MMI-MZI switch with suspended phase arms and heater-on-slab structure," Opt. Lett. **46**, 234-237 (2021).
29. L. Chen and Y.-k. Chen, "Compact, low-loss and low-power 8×8 broadband silicon optical switch," Opt. Express **20**, 18977-18985 (2012).
30. S. Zhao, L. Lu, L. Zhou, D. Li, Z. Guo, and J. Chen, "16 × 16 silicon Mach-Zehnder interferometer switch actuated with waveguide microheaters," Photon. Res. **4**, 202-207 (2016).
31. L. Qiao, W. Tang, and T. Chu, "32 × 32 silicon electro-optic switch with built-in monitors and balanced-status units," Scientific Reports **7**, 42306 (2017).
32. K. Tanizawa, K. Suzuki, M. Toyama, M. Ohtsuka, N. Yokoyama, K. Matsumaro, M. Seki, K. Koshino, T. Sugaya, S. Suda, G. Cong, T. Kimura, K. Ikeda, S. Namiki, and H. Kawashima, "Ultra-compact 32 × 32 strictly-non-blocking Si-wire optical switch with fan-out LGA interposer," Opt. Express **23**, 17599-17606 (2015).
33. L. Song, H. Li, and D. Dai, "Mach–Zehnder silicon-photonic switch with low random phase errors," Opt. Lett. **46**, 78-81 (2021).